\documentclass[conference]{IEEEtran}
\IEEEoverridecommandlockouts
\usepackage{amsmath,amssymb,amsfonts}
\usepackage{algorithmic}
\usepackage{graphicx}
\usepackage{textcomp}
\usepackage{xcolor}
\usepackage{hyperref}
\usepackage{lscape}
\usepackage{pdfpages}
\usepackage[numbers]{natbib}
    
\usepackage{scalerel}
\usepackage{tikz}
\usetikzlibrary{svg.path}

\definecolor{orcidlogocol}{HTML}{A6CE39}
\tikzset{
  orcidlogo/.pic={
    \fill[orcidlogocol] svg{M256,128c0,70.7-57.3,128-128,128C57.3,256,0,198.7,0,128C0,57.3,57.3,0,128,0C198.7,0,256,57.3,256,128z};
    \fill[white] svg{M86.3,186.2H70.9V79.1h15.4v48.4V186.2z}
                 svg{M108.9,79.1h41.6c39.6,0,57,28.3,57,53.6c0,27.5-21.5,53.6-56.8,53.6h-41.8V79.1z M124.3,172.4h24.5c34.9,0,42.9-26.5,42.9-39.7c0-21.5-13.7-39.7-43.7-39.7h-23.7V172.4z}
                 svg{M88.7,56.8c0,5.5-4.5,10.1-10.1,10.1c-5.6,0-10.1-4.6-10.1-10.1c0-5.6,4.5-10.1,10.1-10.1C84.2,46.7,88.7,51.3,88.7,56.8z};
  }
}

\newcommand\orcidicon[1]{\href{https://orcid.org/#1}{\mbox{\scalerel*{
\begin{tikzpicture}[yscale=-1,transform shape]
\pic{orcidlogo};
\end{tikzpicture}
}{|}}}}

\usepackage[normalem]{ulem} 
\usepackage{xcolor}

\usepackage{ifthen}
\usepackage{amssymb}
\newboolean{showcomments}
\setboolean{showcomments}{true} 
\ifthenelse{\boolean{showcomments}}
  {\newcommand{\nb}[2]{
    \fcolorbox{gray}{yellow}{\bfseries\sffamily\scriptsize#1}
    {\sf\small$\blacktriangleright$\textit{#2}$\blacktriangleleft$}
   }
   
  }
  {\newcommand{\nb}[2]{}
   
  }

\usepackage{booktabs}
\usepackage{array}
\usepackage{colortbl}
\usepackage{multirow}
\usepackage{longtable}

\newcolumntype{v}[1]{>{\raggedright \hspace {0pt}}p{#1}}
\newcolumntype{G}[1]{>{\columncolor{gray90}}#1}

\definecolor{Gray}{gray}{0.8}
\definecolor{gray25}{gray}{0.25}
\definecolor{gray50}{gray}{0.50}
\definecolor{gray75}{gray}{0.75}
\definecolor{gray90}{gray}{0.9}

\newcommand{\interviewquote}[2]{\begin{quote}
\footnotesize{\emph{``#1'' }} --- \footnotesize{#2}
\end{quote}}

\def\BibTeX{{\rm B\kern-.05em{\sc i\kern-.025em b}\kern-.08em
    T\kern-.1667em\lower.7ex\hbox{E}\kern-.125emX}}

\begin{document}

\title{

{Requirements Strategy for Managing Human Factors in Automated Vehicle Development}}

\author{\IEEEauthorblockN{Amna Pir Muhammad\IEEEauthorrefmark{1}\IEEEauthorrefmark{2}~\orcidicon{0000-0001-8328-4149}, Alessia Knauss\IEEEauthorrefmark{3}~\orcidicon{0000-0003-4857-7784}, Eric Knauss\IEEEauthorrefmark{1}\IEEEauthorrefmark{2}~\orcidicon{0000-0002-6631-872X}, and
Jonas Bärgmann\IEEEauthorrefmark{1}~\orcidicon{0000-0002-3578-2546}\\
}
\IEEEauthorblockA{\IEEEauthorrefmark{1}\textit{Chalmers University of Technology}} 
\IEEEauthorblockA{\IEEEauthorrefmark{2}
\textit{University of Gothenburg,} }
\IEEEauthorblockA{\IEEEauthorrefmark{3}\textit{Zenseact AB} \\
Gothenburg, Sweden
}
}

\maketitle

\begin{abstract}
%

The integration of human factors (HF) knowledge is crucial when developing safety-critical systems, such as automated vehicles (AVs). 
Ensuring that HF knowledge is considered continuously throughout the AV development process is essential for several reasons, including efficacy, safety, and acceptance of these advanced systems.
However, it is challenging to include HF as requirements in agile development. 
Recently, Requirements Strategies have been suggested to address 
requirements engineering challenges in agile development. 

{By applying the concept of Requirements Strategies as a lens to the investigation of HF requirements in agile development of AVs, this paper arrives at three areas for investigation: a) ownership and responsibility for HF requirements, b) structure of HF requirements and information models, and c) definition of work and feature flows related to HF requirements.}
{Based on} 13 semi-structured interviews with professionals from the global automotive industry, we provide qualitative insights 
{in these three areas}. 
The diverse perspectives and experiences shared by the interviewees provide insightful views {and helped to reason about the potential solution spaces in each area} for integrating HF within the industry, highlighting the real-world practices and strategies used.


\end{abstract}

\begin{IEEEkeywords}
Requirements Engineering, Automated Vehicles, Human Factors, Agile Development, Requirements Strategy
\end{IEEEkeywords}

\section{Introduction}
\label{introduction}



Human factors (HF), a multidisciplinary field focusing on understanding the interactions between humans and other system elements, plays a pivotal role in the design and development of user-centric, safe, and effective automated vehicle (AV) technology \cite{HFJOURNAL}, {including supervised and unsupervised automation. User-centric here means, for example, that the AV technology is
predictable, comfortable, acceptable, and trustworthy.}

\emph{Human factors requirements} refer to those requirements that are primarily introduced due to HF and are essential in this context.
A HF requirement could be that an AV should move to the lane's edge when detecting an oncoming truck, addressing human comfort aspects.

Despite the recognized importance of HF in system development, its integration into the complex and multidimensional environment of AV development \cite{ohnemus1996incorporating,lee2017designing,norman2014things}, especially within agile frameworks \cite{, muhammad2023managing}, remains challenging. 
{For example, it has been reported that} tech leaders and project managers might underestimate the impact of human-related factors, focusing instead solely on technical challenges \cite{zykov2020agile}. 
{In our experience, this oversight is partially due to the technology driven nature of AV development, where the driving task is increasingly perceived as disconnected from humans, focusing instead on computer based automation.
Thus, companies tend to embrace data-driven approaches while also adopting more agile and iterative development paradigms.
However, this may lead to underestimating the role of humans and their capabilities and limitations as constraints for the overall system.}
This underestimation and the inherent complexities of agile methodologies, which emphasize speed and flexibility, often lead to insufficient considerations of HF in the development process {and threatens the safety and success of AV Systems}. 
Recognizing this gap, there arises a critical need for a structured approach to ensure that HF requirements are consistently and effectively integrated. 
This is where the concept of Requirements Strategies comes into play, providing a systematic framework for aligning requirements integration with the dynamic and iterative nature of agile development \cite{muhammad2022defining}.

The Requirements Strategy provides a guideline to integrate requirements engineering practices effectively within agile workflows {\cite{muhammad2022defining}}. 
The core purpose of this framework is to bridge the gap between high-level organizational objectives and the specific requirements engineering activities needed to achieve these objectives. 
This framework facilitates building a shared understanding of the problem space and requirements among all stakeholders involved and to integrate it into a development workflow.
The Requirements Strategy framework consists of three main building blocks: an organizational perspective, a structural perspective, and a work and feature flow perspective. Each of these perspectives contributes to creating a holistic understanding and effective management of requirements.

In this qualitative study, we examine the integration of HF requirements through the lens of Requirements 
{Strategies}. 
Our objective is to evaluate the effectiveness of 
{Requirements Strategies} in facilitating this integration. 
For this purpose, we formulated 
three research questions, each focusing on a main building block of the Requirements Strategy.



\begin{itemize}
\setlength{\itemindent}{0.5em}
\item[RQ1] \textit{Organizational perspective:} How do ownership and responsibility for HF requirements impact their integration in product development?

\item[RQ2] 
\textit{Structural perspective:} How does the structure of requirements and information models impact the integration of HF requirements in product development?

\item[RQ3] 
\textit{Feature and workflow perspective:} How does defining a work and feature flow related to HF requirements influence their integration in product development?

\end{itemize}

For each question, we propose propositions based on common believes 
in existing literature. 
These propositions serve as fundamental hypotheses that guide our analysis, shaping the way we approach and interpret our findings. This method helps us {to} thoroughly examine and apply different aspects of HF 
within the lens of Requirements Strategy.
Using this theoretical framework, we conducted 13 semi-structured interviews with professionals from the global automotive industry. 

Our findings reveal diverse perspectives on the proposed propositions. 
While some cases lend support to our propositions, others suggest that these methods (e.g., requirements decomposition, traceability, assigning HF expertise to critical roles) may not be as effective as anticipated. 
The universal acceptance of these requirements practices as described in the literature and within the framework of Requirements Strategy, is inconclusive in our context. 
These mixed views highlight the necessity for additional research to further explore these findings{, but also the need for organizations to define a consistent Requirements Strategy within the identified solution spaces.}


The remaining paper is organized as follows: 
Section \ref{background} depicts the background 
and 
{anchors our propositions in prior related research}. 
Section \ref{methodology} outlines the research methodology, also addressing 
potential threats to the validity of the results. 
The findings are presented in Section \ref{Findings}. Section \ref{Discussion} engages in a discussion of these findings. Finally, the paper concludes with Section \ref{conclusion }.

\section{Background}
\label{background}

\subsection{Human Factors in Automotive Industry}
{The concept of HF is pivotal in AV development. According to Muhammad et al., ``the field of \emph{Human Factors in AV Development} aims
to inform AV development by providing fundamental knowledge
about human capabilities and limitations throughout the design
cycle so the product will meet specific quality objectives''} \cite{muhammad2023human}{.}

HF experts concentrate on understanding the full spectrum of human attributes, including cognitive, physical, behavioral, psychological, social, emotional, and motivational aspects \cite{HFJOURNAL}. This understanding is vital for designing AVs that are trustworthy, comfortable, effective, and safe {as well as} deliver a satisfying and aesthetically pleasing experience for all stakeholders, as noted in \cite{lee2017designing}.
{The insights into human capabilities and the objectives of AV design are disseminated through various means including design principles, training programs, selection processes, and effective communication strategies. This dissemination is an ongoing necessity, integral to the comprehensive development and refinement of AV systems, throughout the AV design cycle.}

{In Table \ref{tab:examples}, we provide specific examples of functionality of (conditionally) automated vehicles 
together with some important HF aspects for which HF requirements are crucial.
Note that these are examples relating to AV features of different levels of automation that are typically engineering and technology driven, and not to human machine interface (HMI) design. While the design of the latter typically includes HF experts, the former are usually designed and developed in agile teams without integrated HF experts. Also, note that studies have shown a misalignment between engineers' judgments and users' trust in automation \cite{walker2020engineer}.  
}


\begin{table}[tbh]
    \centering
    \caption{{Examples on Human Factors and its impact on AV Design}}
    \label{tab:examples}
    \begin{tabular}{v{0.95\columnwidth}}
\toprule
{\textbf{Lane-Keeping:} Designing and developing a lane-keeping feature presents substantial HF challenges. For instance, automated lane keeping should be designed to minimize the risk of motion sickness \cite{siddiqi2021ergonomic}, while avoiding making interactions with other road-users uncomfortable - for the AV users as well as for other road-users. An example related to user discomfort is that the AV can position itself in the lane in a way that minimizes perceived risk \cite{kolekar2021risk}.
Further, the implemented lane-keeping strategy (e.g., the level of lane-centering) affects the user's acceptance and trust in the system \cite{reagan2017driver}. 
That is, HF requirements include ensuring the lane-keeping algorithm facilitates smooth and predictable movements \cite{tomar2021towards} while incorporating algorithms to maintain safety during close encounters with other vehicles.}
\tabularnewline
\midrule
{\textbf{Lane change:}
 Lane changes must be made in a safe, predictable, comfortable, and acceptable way, for both the users of the AV and for surrounding traffic \cite{jokhio2023analysis}.
Based on user studies, HF requirements may here be operationalized through functional requirements. That is, there may, for example, be specific timings of turn-signal initiations and trajectory shapes that make users (and surrounding traffic) feel that a lane change is predictable and feels comfortable, which in turn facilitates acceptance.}
\tabularnewline
\midrule
{\textbf{Collision Avoidance System Interaction:} In scenarios where the collision avoidance system intervenes to prevent crashes, the timing and nature of the intervention are crucial. For example, nuisance interventions (i.e., when a user does not feel an intervention is warranted) should be kept to a minimum \cite{sander2017opportunities}, while the intervention should be made early enough to avoid crashing. Considering driver comfort zone boundaries \cite{bargman2015quantifying} in collision avoidance design can improve performance while keeping the nuisance intervention rate low.  
Therefore, the collision avoidance algorithm should strike a balance between early intervention (i.e., optimal crash avoidance) and minimizing driver nuisance (e.g., based on driver comfort zone boundaries).
%
Examples of HF requirements in this case include 
seamless interventions 
to maintain passenger trust in the vehicle's safety capabilities, avoiding behaviors that may induce discomfort or distrust.}
\tabularnewline
\midrule
{\textbf{Driver Monitoring System:}
Driver monitoring systems can be used both to mitigate inattention and distraction in lower levels of automation \cite{hobbs1998development}, and to assess if the driver is ready to take over when an AV issues a take-over request (ToR) in higher levels of automation \cite{cirino4640135using}. Both uses are intended to improve safety. For the former, designing warning or intervention strategies that minimize driver nuisance is crucial for acceptance, while it for both is important to consider driver variability and human capabilities.
%
 HF requirements on driver monitoring systems may therefore relate to how and when to provide information to drivers (e.g., warnings or ToRs).}
 \tabularnewline
\bottomrule
    \end{tabular}
\end{table}

Note that this study focuses on HF relevant to the product design and users, not the developers of these systems.

\paragraph{HF Integration in AV Development}
The integration of HF in the automotive industry has emerged as a critical area of focus. Research, as highlighted in studies such as Lee et al. \cite{lee2017designing} and Hancock \cite{hancock2019some}, underlines the vital role of HF in automated system design. This is especially relevant given the increasing complexity of technology and the imperative need for enhanced user safety.
Researchers have studied various key aspects, including the interaction between AVs and human drivers \cite{kyriakidis2019human}, how humans engage with and disengage from these systems \cite{mueller2021addressing}, the design of HMI \cite{ekman2017creating}, and the maintenance of situational awareness \cite{endsley2001designing}. These studies are foundational in comprehending HF's role in elevating the user experience (UX), ensuring safety, and fostering trust and acceptance of AVs.

Despite this, the practical application of HF knowledge often remains limited to specific scenarios or design solutions, usually integrated into the design process or the requirements specification. A gap persists in understanding how HF research insights are effectively translated into the agile development processes of AVs, signaling an area ripe for further exploration and application.

\paragraph{Agile Methodologies and HF Integration}
The challenge of integrating HF into agile development processes in the automotive sector remains a relatively under-explored area, with limited studies specifically addressing this context. 
Saghafian et al. \cite{saghafian2021application} 
{describe} the conflict between the swift and adaptable nature of agile methodologies and the comprehensive, often time-intensive demands of HF practices in  immersive visual technologies field. Steinberg {and Grumman}'s study \cite{steinberg2022human} offers valuable perspectives on modifying agile frameworks to better incorporate HF considerations.
{Muhammad et al. emphasize the role of cross-functional collaboration in successfully integrating HF}
{\cite{muhammad2023human,muhammad2023managing}}. 


These research works highlight the challenges arising from inadequate collaboration and communication among engineering, design, and HF teams, which can hinder the effective integration of HF in development processes. They stress the necessity of establishing robust communication channels and collaborative structures within organizations to facilitate this integration.
However, specific discussions on integrating HF in agile contexts, as outlined in our research, are relatively less explored and represent a contribution to the field.

\subsection{Requirements Strategy and Proposition Formulation}
\label{requiremnets Strategy}
{The concept of a Requirements Strategy allows to reason on a high level of abstraction about how requirements engineering is covered in agile and hybrid software and systems development} \cite{muhammad2022defining}.
{This is particularly applicable since all case companies are agile to some degree; thus, traditional processes do not suffice to express the concepts we need.}
Its primary purpose is to establish a clear and shared understanding of both the problem space and the corresponding requirements. 
A Requirements Strategy comprises three core building blocks: an organizational, 
a structural, 
and a work and feature flow perspective.

\paragraph{Organizational Perspective} This perspective 
in requirements management 
concentrates on defining teams' roles, responsibilities, and the ownership of different types of requirements. 
This aspect is critical for the effective integration and management of requirements in product development \cite{muhammad2022defining}.

A crucial element of requirements management is the clear definition of roles and responsibilities. 
Wiegers and Beatty \cite{wiegers2013software} emphasize the importance of clearly defining roles and responsibilities in the requirements management process.
The significance of assigning clear ownership to each requirement is highlighted by Cagan \cite{cagan2008inspired}, who argues that this clarity fosters accountability and greatly influences project success.
Cockburn \cite{cockburn2006agile} discusses the importance of ownership in agile environments, arguing for a more collaborative and less `siloed' approach.
The literature vividly highlights the importance of role clarity and responsibility in requirements management. This understanding leads to the formulation of our first proposition, emphasizing the impact of clear ownership on HF requirements integration in product development:

\begin{itemize}
   
    \item \emph{Proposition 1:} Clear ownership and responsibility for HF requirements positively impact the integration of HF in product development 

\end{itemize}
Paasivaara and Lassenius \cite{paasivaara2003collaboration} discuss the integration of clear roles and responsibilities within agile methodologies, showing how agile practices can be adapted to improve the management of requirements in dynamic environments. 
Rubin \cite{rubin2012essential} provides insights into the role of product owners in guiding agile teams, hinting to the importance of critical roles in clear ownership in product development. Aurum and Wohlin \cite{aurum2003fundamental} contend that such assignments facilitate better decision-making and accountability, leading to more effective management of requirements and resource allocation, thereby contributing to the overall quality and success of the product.
%
Thus, our literature review indicates that if organizations aim to support managing HF knowledge in a Requirements Strategy from an organizational perspective, 
{HF expertise should be assigned to roles that are critical for the agile development workflow, such as product owners.}
We formulate our Proposition 2 as follows:

\begin{itemize}

    \item \emph{Proposition 2:} Clear ownership and responsibility for HF requirements requires to assign HF expertise to critical roles

\end{itemize}

\paragraph{Structural Perspective} This perspective focuses on defining requirements levels, types, and traceability demands, ensuring clarity and structure in how requirements are approached and documented \cite{muhammad2022defining}.

Hull et al.\cite{hull2005requirements} emphasize the pivotal role of clear and structured requirements for successful project outcomes.
This approach was further refined by Davis \cite{davis1993software}, who differentiated user and system requirements, thus initiating a nuanced understanding of requirements levels. Wiegers and Beatty \cite{wiegers2013software} expanded on this by categorizing requirements into business, user, and software types, offering a comprehensive framework for large-scale project management.
{The emphasis on traceability in the structural perspective aligns with the principles outlined by Gotel and Finkelstein \cite{gotel1994analysis} and Rupp et al. \cite{rupp2009requirements}. They advocate for bi-directional traceability, ensuring that each requirement is linked to its origin as well as the implementation.}
%
%
%
%
Literature underscores the importance of well-structured requirements, emphasis{es} traceability, and requirements decomposition. This {need for} clarity and structure in managing requirements structurally inform{s} our next set of propositions:

\begin{itemize}
\item \emph{Proposition 3:} Clear HF requirements structure positively impacts the integration of HF in product development

\item \emph{Proposition 4:} Clear HF requirements traceability positively impacts the integration of HF in product development

\item \emph{Proposition 5:} Clear decomposition of HF requirements with respect to organizational levels positively impacts the integration of HF in product development

\end{itemize}

\paragraph{Work and Feature Flow Perspective} 
This perspective focuses on defining the lifecycle of requirement types and mapping them to the workflow, ensuring that requirements are effectively integrated into the agile development process \cite{muhammad2022defining}.

{The significance of adaptability and iterative development in managing complex requirements is a cornerstone in this context. Authors such as Van Der Vyver et al. \cite{van2003agile} and Schwaber and Beedle \cite{schwaber2001agile} highlight the importance of being responsive to changing requirements.
Cohn and Ford \cite{cohn2003introducing} further elaborate on this by discussing the dynamics of incorporating requirements into agile processes, setting a foundation for understanding the adaptability of agile frameworks to these changes.}
%
%
%
Schwaber and Beedle 
emphasize the importance of well-defined requirements for the success of agile projects, highlighting the need for a common understanding among stakeholders and interdisciplinary collaboration \cite{schwaber2001agile, highsmith2002agile}. Rubin 
further elaborates on this by presenting a model for managing requirements' lifecycle in agile environments, stressing continuous integration and iterative development as central to agile methodologies \cite{rubin2012essential}.
Moreover, research on project management practices highlights that a clear definition of work streams
improves project outcomes \cite{guide2008guide}.
%
%
%
{The role of review and reflection is highlighted as a critical element in agile development. Sutherland and Schwaber \cite{sutherland2013scrum} emphasize the importance of regular review meetings, such as Sprint Reviews, for assessing progress and integrating evolving requirements. These reviews are evaluative and play a crucial role in adapting and refining requirements based on ongoing feedback. Leffingwell \cite{leffingwell2010agile} extends this notion by advocating for a systematic review strategy to enhance requirement{s} quality and maintain team engagement with evolving changes, reinforcing Schwaber and Beedle's focus on continuous improvement \cite{schwaber2001agile}.

In summary, literature underscores the dynamic nature and lifecycle in agile development.
Additionally, the importance of defining clear work streams and the need for regular reviews and reflections are emphasized. These insights inform our final propositions:

\begin{itemize}

\item \emph{Proposition 6:} A strong lifecycle model of HF requirements positively impacts the integration of HF requirements in product development

\item \emph{Proposition 7:} A clear definition of a HF work stream in relation to other work streams positively impacts the integration of HF requirements in product development

\item \emph{Proposition 8:} A clear plan for review and reflection of HF requirements positively impacts the integration of HF requirements in product development
     
\end{itemize}



\section{Methodology}
\label{methodology}

This study adopts a qualitative research design, utilizing a priori coding to analyze semi-structured interview data. The approach was chosen to systematically categorize and interpret the experiences and insights of professionals in the automotive industry.
Central to our method was the formulation and examination of propositions {(Section \ref{requiremnets Strategy})} for each of our research questions, which guided our data collection and analysis process.

\subsection{Data Collection}

\begin{table*}

\caption{Interviewees’ roles and work experience (Experience level: Low= 0–5 years, Medium=5–10 years, High= More than 10 years)}

\label{tab:Paricipant}
\begin{center}
\begin{tabular}{lv{0.25\linewidth}v{0.1\linewidth}v{0.15\textwidth}v{0.3\textwidth}}
\toprule
\textbf{ID} & \textbf{Role} & \textbf{Experience Level} & \textbf{Supplier/OEM} & {\textbf{Agile Methodologies Usage}}
 \tabularnewline
 \midrule
 01 & Senior Manager & High & OEM & {Some projects adopt agile methodologies, 
 while others follow traditional approaches}
 \tabularnewline
 02 & Technology Leader HF \& Automation & High & Supplier \& OEM & {Some projects adopt agile methodologies, while others follow traditional approaches}
 \tabularnewline
 03 & Senior HF Specialist & High & Supplier \& OEM &  {Using SAFe methodology}
 \tabularnewline
 04 & Engineering Manager & High & OEM & {Using
 agile practices}
 \tabularnewline
 05 & HF expert & Medium & Supplier & {Agile methodologies are employed for software aspects, while hardware components follow a more traditional approach}
 \tabularnewline
  06 & Chief Expert HMI & High & Supplier & {Utilizes agile methodologies variably across projects and business units}
 \tabularnewline
  07 & HF Expert & High & Supplier & {Using agile practices}
 \tabularnewline
 08 & Software Developer & High & OEM & {Fully committed to SAFe methodology} 
 \tabularnewline
 09 & Requirements Engineering Researcher \& Tester & High & OEM & {Operates in a hybrid manner}
  \tabularnewline
  10 & HF Specialist \& Automation and control engineer & Low & OEM & {Embraces agile approaches in their tasks, although the organizational structure remains non-agile}
  \tabularnewline
  11 & AD Safety Specialist &  High & Supplier & {Engages in partial agility, incorporating agile practices, particularly in project execution, but not fully aligned with agile principles in strategic aspects}
  \tabularnewline
  12 & Senior Technical Specialist & High & OEM & {Majority of projects are developed using agile methodologies}
  \tabularnewline
   13 & Global Agile Process Lead & High & Supplier & {Adopted a framework based on SAFe principles}
   \tabularnewline

\bottomrule
\end{tabular}
\end{center}
\end{table*}

\paragraph{Semi-Structured interviews}

Data were collected through 13 semi-structured interviews. This format was specifically chosen to probe the propositions related to each research question.
Each interview lasted approximately 50-60 minutes. Given our intent to include companies from various global locations, virtual interviews were the most feasible approach. Consequently, 11 
interviews were conducted using Microsoft Teams or Zoom video calls, while 2 interviews were conducted in person.
The interviews were recorded with the participants' consent and later transcribed verbatim for analysis.

{Prior to} the interviews, we provided details about the current study objectives, mutual expectations and high-level questions.
Each interview began with an introduction to the authors and an overview of the project, outlining {the} study's purpose and objectives. 
We then started with demographic questions, followed by more targeted questions about the integration of HF in automotive development, directly relating to our propositions.
{Interview questions were grouped in relation to the building blocks of Requirements Strategies, and in each block we asked for an assessment on how well the interviewee experiences their context to perform with respect to these aspects.}
The interview questions used can be found 
\href{https://doi.org/10.5281/zenodo.10570063}{here}. 
{All participants have experience working within agile methodologies to varying degrees. While some have fully embraced agile practices, others have incorporated agile methodologies to a lesser extent.
}

\paragraph{Sampling Strategy}

Our selection criteria {included} respondents with expertise in HF who are employed in the automotive industry.
Specifically, we focused on those experts who have a close working relationship with requirements or process design. {Consequently, all our participants had knowledge of HF.
Given our broader focus on investigating the integration of HF expertise in development processes in general, we did not delve into detailed discussions about specific types of HF. Instead, we relied on participants' overall understanding of HF and their involvement in key aspects such as requirements and design. 
This approach enabled us to capture a broad perspective on the role of HF across various domains within the automotive industry.}

The sampling
included individuals who could provide in-depth and specialized knowledge pertinent to our research objectives - based on the author contacts and by using LinkedIn to screen for potential candidates. 
We reached out to professionals from various automotive companies, selecting those who met our criteria and were willing to participate. This approach ensured rich, qualitative data from a diverse range of experts in the field.

In total, we conducted 13 interviews with professionals from 12 different companies, including industry leaders like Mercedes-Benz, and Volvo. 
Each participant was assigned a unique identifier (ID01 through ID13) to maintain confidentiality. 
{In order to recruit these 13 participants, we contacted 39 suitable candidates, often with several reminders. 
We believe that the low response rate may indicate a particularly high workload among the population of suitable candidates.
We would have preferred more participants to ensure that theoretical saturation was reached but had to stop our recruiting efforts.
From the analysis of our results, we do not believe that additional interviews would significantly change our results (later interviews tend to bring up sentiments that were similarly mentioned before) and report our findings now to enable future research in this important field.}
An overview of these participants, highlighting 
their roles and levels of experience, can be found in Table \ref{tab:Paricipant}. 

    

\subsection{Data Analysis}

The data analysis was structured around our a priori coding framework, 
with three codes per proposition (supports, neutral, rejects). 
These codes were applied to the interview transcripts through line-by-line reading, assigning relevant codes to specific statements.
The initial codes were identified before data collection and were refined as the analysis progressed. 
We adopted Saldana's guidelines for coding as outlined in \cite{saldana2021coding}. The analysis process involved:

\begin{enumerate}
    \item Familiarization with the data through repeated reading of the transcripts
    \item Applying the a priori codes to the data
    \item Ensuring consistency across transcripts and refining the coding framework/guidelines as necessary
    \item Interpretation of the data in the light of our propositions and the broader study context
\end{enumerate}

{The use of a priori coding facilitated a structured and theory-driven analysis.}
This approach not only ensured a systematic examination aligned with the study's goals but also accommodated the emergence of new insights.

{To ensure uniform understanding and application of the codes, two of the authors were involved in coding. 
While the first author coded all data, a second author coded a subset until we were satisfied with the inter-rater agreement.}
{For this purpose, we relied on Cohen's Kappa \cite{kilicc2015kappa} and iterative coding, evaluation, and improvement of coding guidelines until we reached substantial consensus (Kappa above 0.6)}.



\subsection{Research Validity}

To ensure the validity of the study, careful consideration was given to four key aspects: credibility, transferability, dependability, and confirmability.

Credibility, which pertains to the truthfulness and believability of the findings, faced potential threats due to the possibility of participant bias in semi-structured interviews. 
Participants might tailor their responses based on what they perceive the researcher wants to hear, which could lead to skewed data. 
To mitigate this threat, we ensured to have sufficient time to reach an understanding with interviewees and by asking similar questions from different perspectives.

Transferability, or the extent to which the study's findings can be applied to other contexts, was challenged by the unique and specific contexts of the professionals interviewed. Since these individuals might not represent the entire automotive industry, there was a risk of limited generalizability. 
{To mitigate this threat, we collected demographics and additional information about each interviewee's context. 
We use this information in our interpretation of the results and discussion of their implications.}

Dependability, which concerns the stability of the data over time, was threatened by the rapidly evolving nature of the automotive industry. 
As the industry changes, the study's findings might lose relevance. 
To ensure dependability, a detailed documentation of the research process and the decision-making in coding and theme development was maintained. 
This approach not only enhances transparency but also allows the research process to be replicated and critiqued by others. 
Additionally, consistency in data collection and analysis methods was rigorously maintained.

Finally, confirmability, the degree to which the findings are shaped by the respondents and not researcher bias, was a concern. The subjective interpretation of data by researchers could potentially lead to biased conclusions. 
This was addressed by maintaining a reflexive journal throughout the research process, documenting our thoughts, reflections, and potential biases. 
{We also ensured reliability of coding, by leveraging inter-rater agreement for iterative improvement of coding guidelines.}
{Furthermore, there is a threat that interviewees and researchers have different understandings of critical aspects, such as HF. 
We tried to mitigate this threat through careful sampling and by adopting a dialogue-based interview style rather than a strict question-answer style.
This allowed us to identify potential misunderstandings early and to recover during the interview.}
These practices helped in maintaining objectivity and transparency in the research.


\section{Findings}
\label{Findings}

The summary of the interviewees' stances on our propositions, along with a representative quote, can be found 
\href{https://doi.org/10.5281/zenodo.10570063}{here}.
Figure \ref{fig:overview} offers a 
visual overview of the findings.
In the figure, we derive the solution space from our findings below. 

\begin{figure}[ht]
\centering
\includegraphics[width=0.5\textwidth]{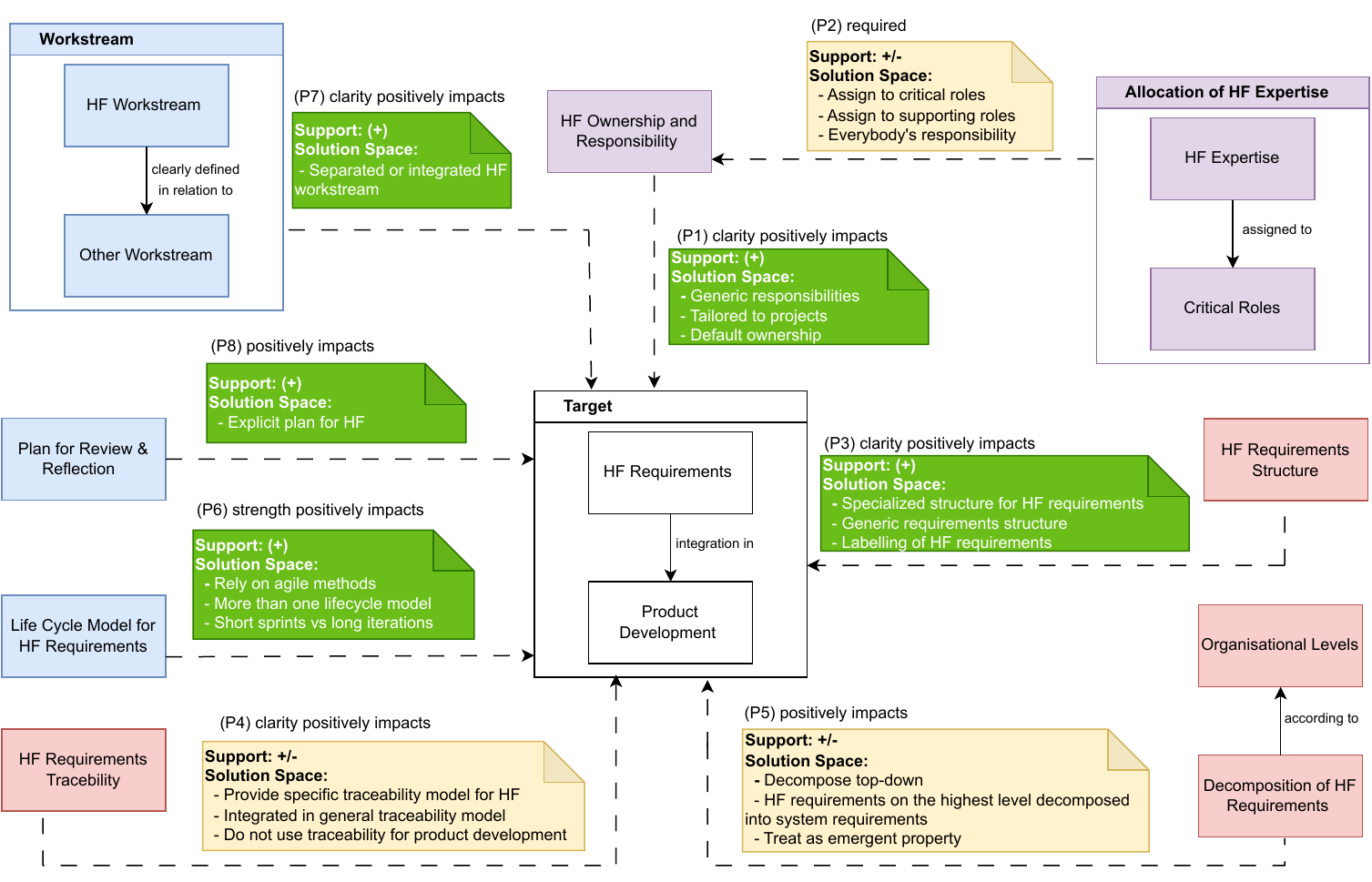}\caption{Overview of findings: Propositions P1 to P8 in relation to RQs (in different colours: RQ1 - purple, RQ2 - red, RQ3 - blue);  propositions for which interviewees expressed a mix of positive and neutral sentiments have green summary boxes, while propositions that also received negative sentiments have yellow boxes.}
\label{fig:overview}
\end{figure}

\subsection{RQ 1: How do ownership and responsibility for HF requirements impact the integration of HF requirements in product development?}
The first research question addresses the organizational aspect of the Requirements Strategy and investigates propositions related to the impact of clearly defined ownership and responsibility on the integration of HF in AV development.

\paragraph{\textbf{Proposition 1: Clear ownership and responsibility for HF requirements positively impact the integration of HF in product development}}

The findings \textbf{support the proposition}, with mostly positive, but also neutral sentiments among interviewees. 
%
%

Many interviewees highlighted defined roles and responsibilities that positively impact the integration of HF in product development, such as the role of a product owner, specialized HF expert or HF team. These roles, as described by interviewees, foster clarity in communication, responsibility allocation, and overall integration of HF into the development cycle. 




For instance, the beneficial influence of assigning HF responsibility to a product owner 
 is illustrated by ID01's experience, where a product owner with UX expertise was able to effectively integrate HF into the development process:

\interviewquote{We hired a product owner who is also a UX person, so she is halftime product owner, half-time UX... So they’re pulling it...the team is basically specifying those requirements.}{ID01}



Similarly, ID04 described a situation where the product owner sets general priorities, while specialists like designers handle detailed aspects of HF. This clear division of responsibility and ownership is exemplified by direct communication between designers and engineers about specific details:

\interviewquote{So the priorities, the overall topics were decided by the {product owner}, but when it came to the nitty-gritty details, it was directly the designer coming to the engineers...}{ID04}

However, several interviewees maintained a neutral stance, discussing various aspects of HF integration and role responsibilities without explicitly supporting or rejecting the proposition. For example, ID05 commented on the integration of HF optimization into developers' responsibilities, indicating a more distributed approach to HF management.

\interviewquote{Optimization with respect to HF is a task that becomes integrated into the developer’s responsibilities.}{ID05}

Overall, the data supports Proposition 1, showing that clear ownership and responsibility for HF requirements positively influence the integration of HF in product development. However, the effectiveness of this integration may vary depending on the organizational structure and specific approaches adopted by different teams or departments.
%

\paragraph{\textbf{Proposition 2: Clear ownership and responsibility for HF requirements requires to assign HF expertise to critical roles}}

Complementary to Proposition 1, we investigate here whether ownership and responsibility should be assigned to critical roles in the development lifecycle,  by making sure that
these roles possess HF expertise. In this context, `critical roles' refer to key positions that significantly influence the direction and success of a project. These roles typically include positions like product owners, portfolio owners, or other decision-making positions. 

The interview data provides a \textbf{nuanced view} on the proposition that assigning HF expertise to {critical roles}
is pivotal for clear ownership and effective management of HF requirements in product development. 
Most interviewees are positive or neutral, but there are two interviewees that take a rejecting stance.

The support to this proposition is evidenced by numerous instances where integrating HF expertise into critical roles enhanced the management and implementation of HF requirements.
%
The interviewees reveal a trend of merging HF expertise with critical roles such as product owners. For instance, ID13 emphasizes the importance of product owners possessing HF knowledge to effectively guide development teams:

\interviewquote{That product owner... should then also
consider... human factors and the human behaviors.}{ID13} 


%
%

%
Similarly, ID08 discusses the responsibilities of portfolio owners:

\interviewquote{The portfolio owner, overseeing a range of products, plays a crucial role at a very high level. Their responsibility is not just to ensure that we meet customer requirements, but also to define and offer products that deliver value to customers.}{ID08}








While designers and engineers can also be considered to be critical roles, we did not count such sentiments as clear support, since the need to have HF expertise covered so broadly shows that, depending on the context, it is more complicated than to just centralize it at selected critical positions.
This interpretation is also supported by other responses that hint towards a
distributed approach to HF responsibility, challenging the notion that clear ownership and responsibilities for HF requirements should be confined to specific critical roles. As for example, mentioned by ID12 and ID05, which contradicts the proposition's focus on assigning HF expertise to specific critical roles.

\interviewquote{The human factors team... they are responsible for all of those elements... including in some cases the design of the actual system itself.}{ID12}
\interviewquote{You cannot really form a group of people, only doing this... Everybody could have some opinion on HF issues.}{ID05}
%


In summary, our data presents a nuanced perspective. While there is substantial support for the proposition that assigning HF expertise to critical roles enhances the management of HF requirements, evidence also points to the effectiveness of a distributed or collaborative approach in certain contexts. This suggests that while assigning HF expertise to critical roles can be beneficial, it may not be the only effective approach for managing HF requirements across different organizational settings.
%


\subsection{RQ 2: How does requirements structure and requirements information model impact the integration of HF requirements in product development?}

The second research question examines how the structuring of HF requirements and the development of information models influence the integration of HF in AV development, assessing the structural part of the Requirements Strategy.

\paragraph{\textbf{Proposition 3: Clear HF requirements structure positively impacts the integration of HF in product development}}

We find this proposition \textbf{rather supported}, with four positive and eight neutral sentiments among interviewees
, but no rejections (one skipped this topic).


Several interviewees highlighted the beneficial impact of a well-defined requirements system on integrating HF in product development. The frequent mention of tools that are used to document requirements and tasks related to requirements for HF such as JIRA, Doors, and Code Beamer underscores their essential role in organizing HF requirements. These tools are key in establishing a structured environment conducive to HF integration.
For example, ID04 emphasizes the effectiveness of a requirements tool for documenting HF aspects, indicating its importance in integrating these factors into product workflows. This highlights the crucial role of structured documentation systems in embedding HF considerations into product development processes.

\interviewquote{In this tool, we documented all the requirements [including design and HF]. We used this as a documentation tool. I'm not sure if that's good or bad, but that's how it was used... I think that the requirements structure worked really well.}{ID04}

Similarly, ID07 described a structured documentation practice for HF studies, indicating a systematic approach that could enhance HF integration in development processes.

\interviewquote{We communicate through these channels, and then we maintain comprehensive documentation of all work that has transpired, including detailed accounts of how studies were conducted and their findings and results.}{ID07}

ID12's shift towards a human-centered design philosophy, with centrally managed requirements, aligns with the proposition, indicating the benefits of a clear, organized structure.

\interviewquote{We have implemented a requirements management system for storing all requirements... [This system is part of] centralizing our design philosophy around being human-centered... everything now stems from this centralized approach...}{ID12}


{In contrast to those who highlight explicit consideration of HF and human-centric considerations when designing the requirements documentation structure,
other interviewees indicate that a generic requirements structure and tool environment can be sufficient or even preferred.
This suggests that specific support for HF considerations is not needed or not beneficial in certain contexts.}

\interviewquote{Partly they are part of the overall documentation environment...there should not be any specific other kind of documentation.}{ID06}

{Moreover}, ID13 speaks to the inclusion of HF within a unified requirements database, lending support to the proposition through a comprehensive approach.
\interviewquote{I would say that it is a part of all other requirements in the requirement database.}{ID13}




Conversely, some responses highlight challenges in the current documentation systems, suggesting that clear requirements structures are not universally effective or consistently utilized.
ID09, for instance, points out a deficiency in HF documentation:
\interviewquote{Human factors requirements are typically not documented explicitly upfront.}{ID09}


ID05 addresses challenges concerning the flexibility and practicality of detailed requirements structuring and documentation systems, emphasizing the extra workload for developers and the difficulty of adapting them in rapidly evolving, innovative settings.
\interviewquote{Documentation is seen as extra work for the developers... everyone deems it as an added task, leading to a general reluctance towards anything additional.}{ID05}


The data presents a composite view. Evidence supports the proposition that a clear requirements structure facilitates the integration of HF into product development, as evidenced by the formal and centralized approaches of some interviewees. 
{This is also indirectly supported by quotes that indicate difficulties when a clear requirements structure is missing.} 
It is however unclear whether this structure must explicitly accommodate HF considerations or should be generic in nature.
The diversity in practices among interviewees suggests that the impact of a clear requirements structure might be influenced by other variables, such as organizational culture or the unique nature of the project.


\paragraph{\textbf{Proposition 4: Clear HF requirements traceability positively impacts the integration of HF in product development}}

The findings indicate a diverse range of practices regarding HF requirements traceability across various organizations. 
This variability reflects differences in both the implementation and the perception of the importance of traceability in integrating HF into product development.


For example, ID01 notes the use of traditional documentation and linking with testing activities, which implies a level of traceability that could support HF integration.

\interviewquote{That describes HF requirements and then from those documents there's some testing... So there's a process for that.}{ID01}

ID06 describes a careful documentation process for high-complexity systems, indicating that in certain contexts, there is clear traceability of requirements, including HF aspects: 

\interviewquote{If we are looking at a, let's say a part  with the compnay's responsibility and higher complexity.....
It [traceability] is extremely carefully taken on all steps of the requirements, it's in the setting...[this includes] clear documentation of possible issues, records of test outcomes,  software version histories, and all these things.}{ID06}




A notable number of interviewees remained neutral. Their responses ranged from acknowledging the use of tools like JIRA for managing requirements to expressing indifference towards the specific traceability of HF requirements. As, for example, ID08 reflects on their organizational structures, which, while not specific to HF, are applied to all requirements. This approach suggests an integration of HF requirements into existing models, but the effectiveness and clarity of these traceability methods are not specified:


\interviewquote{SAFe gives that structure in general, it's not specific for HF, but we applied the same structure for all requirements.}{ID08}

Moreover, ID10 
noted the absence of specific traceability models or processes for HF requirements. This lack of specialized traceability mechanisms could imply a gap in effectively integrating HF in product development.

\interviewquote{ We don't have any traceability whatsoever.}{ID10}

In contrast to the indicators above, some interviewees suggest that traceability does not directly impact the integration of HF in product development. For example, ID05 and ID07 argue that traceability is used for system level reporting and not directly useful for product developers, effectively rejecting our proposition.
\interviewquote{...traceability...more or less only
for reporting, not useful for developing.}{ID05}
\interviewquote{The levels, how you define this is more or less useless for the development.}{ID07}


In summary, the mixed responses from the interviewees illustrate the \textbf{varied perspectives} on the role of traceability in HF requirements integration. While some see clear benefits in traceability for ensuring thorough documentation and process adherence, others view it as having limited practical value in the actual development process. This diversity in opinions suggests that the effectiveness of HF traceability may be highly contextual and dependent on the organizational practices and the nature of the product development processes.


\paragraph{\textbf{Proposition 5: Clear decomposition of HF requirements with respect to organizational levels positively impacts the integration of HF in product development}}

The interview data presents a \textbf{mixed perspective on the Proposition}.

Several interviewees expressed support for the proposition, indicating that decomposing HF requirements across different organizational levels is crucial for their effective integration into product development.



For example, ID12's insights shed light on the decomposition of various requirements under a human-centered design philosophy, which reinforces this proposition. Their approach suggests that systematically addressing HF requirements substantially enhances their integration into product development. 


\interviewquote{At the highest, we go quite deep into the human and understand the requirements of the human and whether they be physical or cognitive and then they are used to guide specific requirements which are then decomposed into the engineering requirements that are used to deliver the car....So the typical and hierarchical level approach}{ID12}




Similarly, ID13’s emphasis on iterative design and user stories further aligns with this perspective, indicating that decomposing HF requirements into manageable parts facilitates improved understanding and implementation:

\interviewquote{Then we can quickly move forward and break that down into the different level of requirements. Implement, test, and validate and come back to you in two or three weeks.}{ID13}

ID09 details a clear multi-level decomposition process, reinforcing the proposition:

\interviewquote{We treat our initial requirements as level one. These are then further decomposed into level two, or system requirements, which may differ from the conventional use of the term. Sometimes, there’s even a third or fourth level for specific needs like FPGA. Generally, we operate with these three levels}{ID09}

However, ID05 and ID08 take a rejecting stance and noted challenges or alternative perspectives. ID05, for example, points out the complexities of decomposing HF requirements in innovative fields like autonomous driving:

\interviewquote{In at least in the new area, in the autonomous driving area... You cannot really decompose the requirement at the beginning because you don't know what the technique can do, what the sensor can see, what the system is capable of.}{ID05}

ID08’s observations suggest variability in the application of structured decomposition across different organizations:

\interviewquote{I would say we don't have the structure as you would probably want to see it...At the moment we don't have a standardized taxonomy or something like that.}{ID08}

Some interviewees remained neutral, neither supporting nor rejecting the proposition outright. As, for example, ID07's comment implies a more adaptive rather than structured approach to HF integration.

\interviewquote{We sort of adapt to it and then we try to decompose according to the order}{ID07}


In summary, the mixed responses highlight the complexity and contextual nature of integrating HF requirements in product development. While a structured decomposition approach is favored by some, the diversity in organizational practices and the abstract nature of HF requirements pose challenges to its universal application. These findings suggest that while the proposition may hold value, its applicability and effectiveness are likely contingent on the specific context of the organization and the nature of the product development process.

\subsection{RQ 3: How does defining a work and feature flow related to HF requirements influence their integration in product development?}
%
This question evaluates the third aspect of requirement strategy, i.e., the work and feature flow perspective.

\paragraph{\textbf{Proposition 6: A strong lifecycle model of HF requirements positively impact the integration of HF requirements in product development}}

Our findings \textbf{support Proposition 6} (7 positive, 5 neutral stances among interviewees), particularly highlighting the benefits of integrating agile practices.
%
Our analysis reveals a trend towards the integration of HF requirements within agile and iterative development processes, reinforcing the notion that a robust lifecycle model of HF requirements is instrumental in enhancing product development. 
For example, ID01 provided insight into their organization's structured approach, illustrating the systematic adoption of agile methodologies:
 
\interviewquote{We're doing we have the set of meetings, there's the planning...understanding requirements, implementing, testing, and integration., get feedback and iterate based on that.}{ID01}

This approach underscores the efficacy of agile methodologies as integral components of the lifecycle model, particularly in their role in facilitating the integration of HF requirements.
 The iterative nature of agile methodologies, as pointed out by our interviewees, enables the seamless incorporation of various specialties, including HF. ID13 elaborates on this process:

\interviewquote{In an agile way of working, you're working incrementally and iteratively. So we're doing one functional feature at a time, we use different modeling and simulation before we have the hardware available to us...you're doing some of the HF requirements depending on what you're developing next. [\ldots] Well, it works very well with incrementally in an iterative approach.}{ID13}

The emphasis on early decision-making and continuous feedback loops, as observed by ID12, further corroborates the value of a well-structured lifecycle model for HF integration:

\interviewquote{Lifecycle model involves early-stage requirements capture, evaluation, and iterative feedback... Separates design and delivery aspects....Significant architectural decisions, some of the big decisions that really have to be defined early on and not changed and made with always people being involved from this part of the process.}{ID12}

However, some interviewees expressed neutral stances, indicating variability in the application of lifecycle models or the complexity of the project. For instance, ID10 noted:

\interviewquote{The process isn’t as intuitive or straight- forward. It often requires a significant amount of repeated effort, depending on the complexity of the issue this iterative process involves a few steps for simpler problems but becomes more detailed when tackling complex issues.}{ID10}

In summary, the data support the proposition that a robust lifecycle model for HF requirements positively influences their integration in product development, as evidenced by the majority of interviewees. However, the neutral responses indicate variability and complexity in the application of these models, suggesting that while a strong lifecycle model is generally beneficial, its implementation and effects may vary depending on specific project contexts and methodologies.

\paragraph{\textbf{Proposition 7: A clear definition of a HF work stream in relation to other work streams positively impacts the integration of HF requirements in product development}}

Our findings provide \textbf{partial support for Proposition 7}, as three interviewees supported the proposition while eight others maintained a neutral stance.

Some interviewees suggest that when HF is clearly integrated into the workflow, particularly in agile environments, it is more effectively considered in product development.
For instance, ID02 emphasizes that agile methodologies facilitate the integration of various specialists, including HF specialists. This is evident from the observation,

\interviewquote{Human factors specialist and HF requirements should be part of the regular work, together with everything else, we should not keep it separate because that's what we have today. Today we are in separate, we have this, these silos. We want to break silos with agile way of working.}{ID02}

This statement supports the proposition, as it demonstrates that clear and integrated work streams, such as those found in agile environments, are conducive to incorporating HF requirements effectively. Similarly, ID13's perspective reinforces this view by advocating for a unified approach where HF is integrated along with other critical functions:

\interviewquote{There are no separate work streams for HF. The aspects are all the same, we don't have separate works streams for functional safety or for cyber security or anything else like that.}{ID13}

However, some interviewees expressed a neutral stance, indicating a diverse implementation of HF integration in their processes.
For instance, ID04's comment reveals a gap between intention and execution:

\interviewquote{The developers implementing the code always wanted to have the human factor requirements... the designers did not work with our PI planning.}{ID04}

ID03's experience further illustrates the challenges of integrating HF into other workstreams, hinting at possible technical complexities:

\interviewquote{It's integrated, but sometimes I feel it's very technical.}{ID03}

Moreover, we learned that the impact of clearly defined HF work streams on their integration appears to be context-dependent, varying across different organizational structures and project types. ID12's experience illustrates this variability:

\interviewquote{In some cases, separate teams for HF run their own sprints, typically delivering specifications because we don't need external resources. Alternatively, people work with joint teams to develop requirements that consider all aspects, including mechanical, electrical, and others.}{ID12}

In summary, while the data provides evidence that clear definitions and structured work streams in agile methodologies can improve the integration of HF in product development, there is also an understanding that its effectiveness can vary based on project dynamics and organizational frameworks.  
Notably, none of the interviewees explicitly rejected the proposition.

\paragraph{\textbf{Proposition 8: A clear plan for review and reflection of HF requirements positively impacts the integration of HF requirements in product development}}
Our findings  
\textbf{support Proposition 8}, with seven interviewees taking a supporting stance and five others taking neutral stances.
%
%
Our interviewees highlighted practices and perspectives that align with the importance of a structured approach to reviewing and reflecting on HF requirements. This support is evidenced by the emphasis on agile flexibility, the importance of planning and communication, and the iterative improvement processes discussed by various interviewees. 
Iterative processes and continuous improvement, central to agile methodologies, are also highlighted as beneficial for integrating HF. For instance, ID12's emphasis on retrospectives underscores this point:

\interviewquote{We'll conduct retrospectives and discuss the process, people, and output. The main feedback, however, is whether what we've created works, is usable, and if people like it. Does it feel intuitive?... Continuous improvement is key. At the end of the day, it's about spending time, doing the work, reviewing, reflecting, and improving for the next time.}{ID12} 

Similarly, ID13's mention of sprint reviews and the role of retrospectives as a platform for stakeholder engagement to reflect on HF requirements indicates a structured approach:

\interviewquote{The Sprint review is the time and place for the team to invite stakeholders to review and reflect on the human factor requirements. As I said, this should preferably be done with the stakeholders. The team should work as closely as possible with them, regardless of who the stakeholders are.}{ID13}

These insights illustrate the positive impact of regular review and reflection in aligning product development with HF requirements. ID07's comments further support this, illustrating a formalized review process:

\interviewquote{When you complete a study or project, you usually create an HF requirements report deliverable. It's reviewed by the authors or peer-reviewed, and then you receive feedback, addressing minor or major comments. We also typically have lessons learned, discussing improvements for future projects.}{ID07}



%
{Some interviewees did not explicitly support or reject the proposition. As for example, ID09 highlights the importance of the review and reflection processes but does not directly link these to the positive integration of HF requirements.}
\interviewquote{We review and reflect overall for the workflow and testing.}{ID09}

{In summary,  the data predominantly supports the proposition that a clear plan for review and reflection of HF requirements is beneficial for their integration in product development. The emphasis on iterative feedback, stakeholder involvement, and retrospective analyses by several interviewees underscores the value of reflective practices in effective product development. However, the presence of neutral stances suggests that the impact of such practices may vary depending on other factors, such as team dynamics, project nature, or organizational contexts.}

\section{Discussion}
\label{Discussion}


Our findings have been obtained from a qualitative analysis of semi-structured interview data based on the lens of Requirements Strategies.
In this section, we discuss how these findings are contextualized within existing literature, drawing parallels and contrasts with strategies in related fields.
{Our findings indicate that while every company adopts agility to varying extents, they do so in distinct ways tailored to their specific contexts. Consequently, these results may have implications for organizations that go beyond `traditional' processes at both system and organizational levels.}

 \paragraph{Impact of Ownership and Responsibility on HF Integration}
The results indicate general support confirming that clear ownership and responsibility (Proposition 1) benefit HF integration in product development, resonating with Smith and Reinertsen’s \cite{smith1997developing} observations on role clarity.  However, the neutral stances suggest a nuanced application, dependent on organizational context and project nature. This aligns with the findings of Cockburn \cite{cockburn2006agile}, who highlight the impact of organizational culture on the adoption of agile practices.
For Proposition 2, the opinions on assigning HF expertise to critical roles were divergent. 
Hence, we conclude that this is not the sole effective approach. 
%
The diversity in responses implies that either a distributed approach to managing HF expertise or that a combined approach that distributes HF expertise generally over the development organization, and at the same time ensures that critical roles can utilize it in their decision making can be effective.

\paragraph{Impact of Requirement Structuring and Information Models on HF Integration in AV Development}


The mixed perspectives on the impact of clear requirements structure on HF integration (Proposition 3) 
{echo} the broader discourse in software engineering \cite{sommerville2011software}, where structured approaches, while valuable, must be flexible enough to accommodate the dynamic nature of product development, as noted by Cooper et al. \cite{cooper2014face}.
This also reflects the tension in software development 
expressed by the controversial agile value of working software over comprehensive documentation \cite{davis2013just, beck2001manifesto}. 
{Our findings contradict literature such as Hull et al. that describes the clarity and structure of requirements as pivotal for successful project outcomes \cite{hull2005requirements}, at least in the context of our investigation. 
We believe that due to the disruptive nature of AV technology, some of our interviewees are suspicious about a structure that may prevent them to explore emergent HF properties.
The fact that not all areas of automotive development are equally disrupted may explain why our interviewees come to different assessments.}

For Proposition 4, regarding traceability of HF requirements, the diversity of viewpoints underscores the practical challenges of maintaining traceability in various organizational contexts. This reveals that the effectiveness of traceability in HF integration may not be universally recognized, suggesting a gap between theoretical expectations \cite{rupp2009requirements, gotel1994analysis}and practical realizations.
%
The findings on the decomposition of HF requirements in Proposition 5 show varied support.
{The iterative approaches to requirement decomposition discussed by interviewees reflect systems engineering best practices for breaking down complex requirements into manageable parts for efficient implementation. }
{However, other interviewees felt that requirements decomposition does not work well in innovative domains, where traditional decomposition schemes may not fit well.
Similar effects have also been reported in literature, e.g. by Hoda et al. \cite{hoda2018rise}.}

\paragraph{Impact of Work and Feature Flow on HF Integration in AV Development}

The findings regarding the lifecycle model of HF requirements (Proposition 6) support the positive impact of the integration of HF in product development. This aligns with agile methodologies which emphasize adaptability and iterative development for complex requirements, as discussed in Van Der Vyver's et al. \cite{van2003agile} and Schwaber and Beedle's \cite{schwaber2001agile} work.
Our study shows both positive and neutral sentiments among interviewees, which suggests diverse approaches to integrate HF into the development, 
influenced by factors like project size, nature, organizational culture, and scope \cite{sutherland2007scrum}.
%
%

The feedback on Proposition 7 highlight the debate on integrating versus separating HF activities in agile processes, echoing Cockburn and Highsmith's view on agile methodologies' adaptability across organizational contexts \cite{cockburn2001agile}.
Integrating HF with other workstreams may improve recognition and integration, but might also limit its scope to specific areas. Conversely, external HF expertise may have a broader reach across teams, but have less influence on decision-making. 
This balance between specialized knowledge and cross-functional collaboration is also discussed by Highsmith \cite{highsmith2002agile}.

Lastly, the emphasis of Proposition 8
on the importance of a clear plan for review and reflection of HF requirements is supported by Sutherland and Schwaber's insights \cite{sutherland2013scrum} into continuous improvement and iterative evaluation in agile development.
The mix of positive and neutral responses indicates differences in how organizations implement these processes, ranging from specialized review approaches to relying on generic reviews schemes. 
This divergence likely stems from their distinct cultural foundations. For example, in firms where a human-centric design is deeply embedded, additional reporting may be viewed as redundant.
However, we believe that in companies still developing this culture, detailed reports could be crucial for increasing awareness and driving cultural change.


Overall, the lens of Requirements Strategies allowed for meaningful discussions with experts,
{particularly benefiting companies transitioning away from purely process-driven approaches.}
The fact that we find a wide spectrum of different approaches for each proposition, and even a certain level of disagreements with the structural propositions, shows that a suitable Requirements Strategy would have to be defined for each specific context. 
For this reason, we derive different solutions from our interviews, to indicate the solution space for organizations that aim to integrate HF requirements into AV development.
We believe that our findings will help organizations to explore this solution space and make good, consistent decisions. 
We are certain that this will help to systematically cover HF concerns in AV development.

\section{Conclusion}
\label{conclusion }

{In conclusion, this study contributes to the understanding of how HF requirements can be effectively integrated into AV development, examined through the lens of Requirements Strategies. 
Adopting agile methodologies has proven promising in addressing complex HF requirements through iterations and feedback loops, yet it becomes unclear how HF requirements can be systematically managed in agile scopes.
This tension highlights the crucial role of a consistent Requirements Strategy in the successful integration of HF knowledge into the agile development of AVs. 
Our findings underscore the necessity for organizations to adapt their strategies to the unique demands of the development environments.}
In particular, they indicate that clear ownership and responsibility for HF requirements 
enhance their integration into product development, resulting in more streamlined and cohesive workflows.
While foundational, the structuring and traceability of HF requirements also necessitate a balance between rigidity and flexibility. 
Both, organizational matters and structuring of requirements are fundamental to integrate activities related to HF requirements into agile workflows.
{Along these three components (organization, structure, workflow), our study identifies various solution spaces indicated in the findings, reflecting the diverse approaches employed within the industry to integrate HF requirements. 
We observed that while certain practices are effective, the effectiveness of others
may vary, highlighting the need for future research and the development of tailored approaches that consider the unique organizational contexts.}


\section*{Acknowledgments}
The authors extend their sincere gratitude to all 
interviewees 
for their 
time and valuable insights. 
This work was supported by funding from the European Union’s Horizon 2020 research and innovation programme, through the Marie Skłodowska-Curie grant, agreement number 860410. For additional information about this project, please visit https://www.shape-it.eu.

\bibliographystyle{elsarticle-num}
\bibliography{bib}



\end{document}